\documentclass[12pt,preprint]{aastex63}

\usepackage[caption=false]{subfig}
\usepackage{multirow}
\usepackage{graphicx}
\usepackage{booktabs}
\usepackage{amsmath,amssymb}
\usepackage[T1]{fontenc}

\graphicspath{{./}{figures/}}

\begin{document}

\title{Nearby SN-Associated GRB~190829A: Environment, Jet Structure, and VHE Gamma-Ray Afterglows}
\correspondingauthor{En-Wei Liang}
\email{lew@gxu.edu.cn}
\author{Lu-Lu Zhang}
\affil{Guangxi Key Laboratory for Relativistic Astrophysics, Department of Physics, Guangxi University, Nanning 530004, China;lew@gxu.edu.cn}
\author{Jia Ren}
\affil{School of Astronomy and Space Science, Nanjing University, Nanjing 210023, China}
\author{Xiao-Li Huang}
\affil{School of Astronomy and Space Science, Nanjing University, Nanjing 210023, China}
\author{Yun-Feng Liang}
\affil{Guangxi Key Laboratory for Relativistic Astrophysics, Department of Physics, Guangxi University, Nanning 530004, China;lew@gxu.edu.cn}
\author{Da-Bin Lin}
\affil{Guangxi Key Laboratory for Relativistic Astrophysics, Department of Physics, Guangxi University, Nanning 530004, China;lew@gxu.edu.cn}
\author{En-Wei Liang*}
\affil{Guangxi Key Laboratory for Relativistic Astrophysics, Department of Physics, Guangxi University, Nanning 530004, China;lew@gxu.edu.cn}

\begin{abstract}
We present a self-consistent paradigm for interpreting the striking features of nearby low-luminosity GRB~190829A. Its prompt gamma-ray lightcurve has two separated pulses. We propose that the interaction of the hard prompt gamma-ray
photons ($E_p= 624_{-303}^{+2432}$ keV) of its initial pulse with the dusty medium ($A_{\rm V}=2.33$) does not only result in the second soft gamma-ray pulse ($E_p\sim 12$ keV),
but also makes a pre-accelerated $e^{\pm}$-rich medium shell via the $\gamma\gamma$ annihilation.
In this paradigm, we show that the observed radio, optical, and X-ray afterglow lightcurves are well fit with the forward shock model. Its jet is almost isotropic ($\theta_j>1.0$ rad) with a Lorentz factor of $\sim 35$, and the electron density of the $e^{\pm}$-rich medium shell is $\sim 15$ cm$^{-3}$,
about 7~times higher than the electron density of its normal surrounding medium. The GRB ejecta catches up with and propagates into the $e^{\pm}$-rich medium shell
at a region of $R=(4.07-6.46)\times 10^{16}~\rm cm$, resulting in a bright afterglow bump at $\sim 10^3$ seconds post the GRB trigger. The predicted very high energy (VHE) gamma-ray emission from the synchrotron self-Compton process agrees with the H.E.S.S. observation. The derived broadband spectral energy distribution shows that GRB~190829A like nearby GRBs would be promising targets of the VHE gamma-ray telescopes, such as H.E.S.S., MAGIC, and CTA (Cherenkov Telescope Arrays).
\end{abstract}

\keywords{gamma-ray burst: general - gamma-ray burst: individual (GRB 190829A)}

\section{Introduction}
\label{sec:introduction}
As the most intense burst phenomena in the universe,
gamma-ray bursts (GRBs) and their afterglows are theoretically predicted
as sources of  very high energy (VHE) gamma-rays and cosmic rays \citep{Waxman-1995-PhRvL..75..386W,Milgrom-1995-Usov-ApJ...449L..37M,Vietri-1995-ApJ...453..883V,
Abdalla-2019-Adam-Natur.575..464A,Samuelsson-2020-Begue-ApJ...902..148S}.
They are listed as the target sources of current and future telescopes in the GeV-TeV gamma-ray bands.
It is exciting that sub-TeV gamma-ray emission was firstly convincingly detected in the early
afterglows of GRB~190114C with the Major Atmospheric Gamma Imaging Cerenkov Telescopes (MAGIC; \citealp{MAGIC_Collaboration-2019-Acciari-Natur.575..455M}).
Its broad-band spectral energy distribution (SED) in the optical, X-ray,
and gamma-ray bands can be well explained with the models of synchrotron
radiation and synchrotron self-Compton (SSC) process of the electrons accelerated in the jet \citep{MAGIC_Collaboration-2019-Acciari-Natur.575..459M,Derishev-2019-Piran-ApJ...880L..27D,Wang-2019-Liu-ApJ...884..117W}.
The SSC component is also marginally detected in GRB~130427A
\citep{Liu-2013-Wang-ApJ...773L..20L,Ackermann-2014-Ajello-Sci...343...42A,
Joshi-2019-Razzaque-arXiv191101558J,Huang-2020-Liang-ApJ...903L..26H}
and GRB~180720B
\citep{Fraija-2019-Dichiara-ApJ...885...29F,Wang-2019-Liu-ApJ...884..117W,Duan-2019-Wang-ApJ...884...61D}.
The window in the VHE gamma-ray band is opened for the GRB study.
This is not only benefical to reveal the radiation physics,
but also critical for exploring the burst environment (e.g., \citealp{Huang-2020-Wang-arXiv201213313H}).

Typical GRBs are happened at the high-redshift universe \citep{Salvaterra-2015-JHEAp...7...35S}.
VHE photons in the TeV-band are suffered sharply absorbed by the extragalactic background light (EBL)
via the electron pair production (e.g., \citealp{Stanev-1998-Franceschini-ApJ...494L.159S,
Dwek-2005-Krennrich-ApJ...634..155D,Dwek-2005-Krennrich-ApJ...618..657D, Aharonian-2006-Akhperjanian-Natur.440.1018A,Mazin-2007-Raue-A&A...471..439M,
Franceschini-2008-Rodighiero-A&A...487..837F,Meyer-2012-Raue-A&A...542A..59M,
Abdollahi-2018-Ackermann-Sci...362.1031F}).
This is a great obstacle for the detection of VHE gamma-rays from typical GRBs.
The detection of the VHE gamma-ray afterglow of GRB~190114C is some lucky since it is among
the most energetic burst happened at a relatively low redshift ($z=0.4245\pm0.0005$; \citealp{Castro-Tirado-2019-Hu-GCN.23708....1C}).
Note that nearby low-luminosity GRBs (LL-GRBs) seem to be a unique GRB population
that is characterized by a high local event rate and large jet opening angle (e.g., \citealp{Liang-2007-Zhang-ApJ...670..565L}).
It was proposed that they are considerable sources of VHE photons and cosmic rays \citep{Murase-2019-ICRC...36..965M}.

Interestingly, GRB~190829A is a LL-GRB associated with a broad-line type Ic supernova,
SN~2019oyw \citep{de_Ugarte_Postigo-2019-Izzo-GCN.25677....1D,Hu-2020-Castro-Tirado-arXiv200904021H}.
It is among the nearest GRBs with a redshift of $z=0.0785\pm0.005$ \citep{Valeev-2019-Castro-Tirado-GCN.25565....1V}. Its prompt gamma-rays have two distinct pulses, i.e., a hard-weak pulse followed by a soft-bright pulse with a separation of about 50 seconds \citep{Chand-2020-Banerjee-ApJ...898...42C},
but the two pulses do not share the same $L_{\rm iso}-E_p$ relation as usually seen in long GRBs \citep{Lu-2012-Wei-ApJ...756..112L}.
Its high energy afterglow was detected by H.E.S.S. at 4.3 hours after the GRB trigger \citep{de_Naurois-2019-H.E.S.S._Collaboration-GCN.25566....1D}.
\cite{Zhang-2020-Murase-arXiv201207796Z} suggested that the VHE
gamma-ray afterglow is produced in the external inverse-Compton scenario for
seed photons supplied by the second emission episode of the prompt gamma-rays,
but the bump have a long delay to the prompt gamma-rays.
The optical and X-ray light curves of GRB~190829A afterglow have an achromatic bump with a rapid increase
at around $700$ seconds post the GRB trigger,
and the X-ray afterglow faded as a single power-law up to more than 110 days
without a detection of the jet break.
The origin of the bump is uncertain. It was proposed that the bump may be
due to the activity of the GRB central engine \citep{Chand-2020-Banerjee-ApJ...898...42C}
or the dipole radiation of its remnant magnetar plus the forward shock emission \citep{Fraija-2020-Veres-arXiv200311252F}.
Note that, similar optical afterglow bump was also observed in some GRBs,
e.g., GRB~970508 \citep{Castro-Tirado-1999-Gorosabel-A&AS..138..449C}.
\cite{Dai-2002-Lu-ApJ...565L..87D} proposed that the bump is a signature of medium density jump.
The radio afterglow of GRB~190829A was detected at an even later epoch than the X-ray afterglow and lasted longer \citep{Rhodes-2020-van_der_Horst-MNRAS.496.3326R}.
The interpretation of the radio afterglow is also debated.
\cite{Rhodes-2020-van_der_Horst-MNRAS.496.3326R} suggested that the radio afterglow in 15.5~GHz and 1.3~GHz
is from the reverse and forward shocks, respectively.
\cite{Sato-2021-Obayashi-arXiv210110581S} proposed that the observed low luminosity of
GRB~190829A is due to the off-axis observation of a two-component jet and late X-ray and radio emissions are from the wide jet component.

In this paper, we revisit the data of GRB 190829A and present a self-consistent
interpretation of its features  mentioned above.
We find that the radio, optical, X-ray afterglow data are well fitted with the forward shock model,
and the VHE gamma-ray afterglow is resulted from the SSC process of the electrons in the jet
without inducing an external seed photon field as proposed by \cite{Zhang-2020-Murase-arXiv201207796Z}.
Our data analysis is present in \S~{\ref{Data Analysis}},
and the afterglow light curve modeling is reported in \S~{\ref{Afterglow Modeling}}.
We discuss the VHE emission of GRB~190829A afterglow in \S~{\ref{VHE_afterglow}}.
Conclusions and discussion on our results are presented in \S~{\ref{Conclusions}}.

\section{Data Analysis}
\label{Data Analysis}
GRB~190829A was detected by {\em Fermi}/GBM, {\em Swift}/BAT, and {\em Konus}/Wind satellites \citep{Lesage-2019-Poolakkil-GCN.25575....1L,
Lien-2019-Barthelmy-GCN.25579....1L,Tsvetkova-2019-Golenetskii-GCN.25660....1T}.
Figure \ref{Prompt_LC} shows its BAT light curve.
It has two separated pulses.
The spectrum of the first pulse in the 20 keV - 2 MeV range observed with Konus/Wind is well fitted with
a power-law with exponential-cutoff model,
yielding a photon index of $\alpha = -1.33^{+0.30}_{-0.23}$
and the peak photon energy of the $\nu f_\nu$ spectrum of $E_p= 624_{-303}^{+2432}$ keV
in the burst rest frame \citep{Tsvetkova-2019-Golenetskii-GCN.25660....1T}.
The second pulse is very soft and its spectrum observed with {\em Fermi}/GBM in the 8-1000 keV band
can be fitted with a Band function \citep{Band-1993-Matteson-ApJ...413..281B}
with $E_p=11.8\pm 1.1$ keV in the burst rest frame,
$\alpha=-0.92\pm 0.62$, and $\beta = -2.51\pm 0.01$ \citep{Lesage-2019-Poolakkil-GCN.25575....1L}.

We collect the afterglow data from the literature and GCN reports,
and their light curves are shown in the left panel of  Figure~{\ref{multiwavelength}}.
Note that the optical light curve is shown with the data in u band taken from
\cite{Chand-2020-Banerjee-ApJ...898...42C} and \cite{Hu-2020-Castro-Tirado-arXiv200904021H}
since the data have good temporal coverage.
The optical data has been corrected by extinctions from our Galaxy
and the GRB host galaxy as discussed below.
One can observe that the X-ray and optical light curves show the same behavior.
They initially keep almost a constant and rapidly increase since
$t>700$ seconds and reach the peak at around $t=1200$ seconds.
The fluxes decay as normal afterglow from external shocks up to
$\sim 10^{7}$ seconds in the X-ray band without the detection of jet break.
Radio afterglows observed with the Meer Karoo Array Telescope (MeerKAT, 1.3~GHz, \citealp{Monageng-2019-van_der_Horst-GCN.25635....1M}) and
Arcminute Microkelvin Imager-Large Array (AMI-LA, $15.5$~GHz) were detected from around one day
to about 200 days post the GRB trigger \citep{Rhodes-2020-van_der_Horst-MNRAS.496.3326R}.
Note that the fluxes in the X-ray and 15.5 GHz band at $t>5.5\times 10^{6}$ seconds keep as constant, i.e.,
$F_{\rm X}=1.68\times 10^{-13}$ erg cm$^{-2}$ s$^{-1}$ and $F_{\rm 15.5GHz}=150~\rm \mu Jy$.
They should be contributed by the GRB host galaxy.
We subtract them from the X-ray flux and radio flux in the 15.5~GHz in our fitting of the light curves.

We make joint spectral analysis for the optical and X-ray band afterglows in the time interval $[3,4]\times10^{4}$ seconds.
The X-ray data are observed with the {\em Swift}/XRT\footnote{Collected from \url {https://www.swift.ac.uk/xrt_spectra/}.}.
The quasi-simultaneous optical $g, r, i, z, \rm J, H, Ks$ band data are taken from \citep{Chen-2019-Bolmer-GCN.25569....1C}
and $u$ band data is taken from \cite{Hu-2020-Castro-Tirado-arXiv200904021H},
as tabulated in their Table~3.
All the optical data are corrected by the Galactic extinction \citep{Schlafly-2011-Finkbeiner-ApJ...737..103S},
and the Galactic HI column density is fixed at $N_{\rm H}^{\rm Gal}= 5.6\times 10^{20}~{\rm cm}^{-2}$. The extinction-law of the GRB host galaxy is taken the same as our Galaxy, i.e., $R_{\rm V}=3.08$ \citep{Pei-1992-ApJ...395..130P}. An absorbed power-law model is equated to fit the spectrum,
yielding a HI column density of the GRB host galaxy at $z=0.0785$ as
$N_{\rm H}^{\rm host}=(5.79\pm0.53)\times10^{21}~{\rm cm}^{-2}$, the optimal color index as $E_{\rm B-V}=0.76\pm 0.01$ corresponding to $A_{\rm V}=R_{\rm V}E_{\rm B-V}=2.33$. The photon index is $1.76\pm 0.01$. Our fit is shown in the right panel of Figure~{\ref{multiwavelength}}.
Strong extinction indicates that the burst environment is massively dusty. The u band data shown in Figure \ref{multiwavelength} are corrected with $A_{u^{\prime}}=1.579$ \citep{Schlegel-1998-Finkbeiner-ApJ...500..525S}.

\section{A Self-Consistent Paradigm and Multiwavelength Afterglow Modeling}
\label{Afterglow Modeling}
Basing on the data analysis above,
we propose a paradigm to explain both the prompt and afterglow emission of GRB~190829A.
We outline the paradigm as following.
First, the ejecta powered by the central engine of GRB 190829A is quasi-isotropic
since no jet break is detected later than 110 days post the GRB trigger.
Second, the initial hard gamma-ray pulse is produced by internal shocks of the ejecta.
Third, the hard gamma-ray photons ($E_p= 624_{-303}^{+2432}$ keV in the burst rest frame)
are scattered by the density medium $(A_V=2.33)$,
resulting the second soft gamma-ray pulse ($E_p=11.8\pm 1.1$ keV in the burst rest frame),
similar to that proposed by \cite{Shao-2007-Dai-ApJ...660.1319S}.
Forth, the interaction of the gamma-ray photons with the dust
leads to an $e^{\pm}$-rich medium shell via the $\gamma\gamma$ annihilation,
and the $e^{\pm}$-rich medium shell is also pre-accelerated by the gamma-ray photons
\citep{Madau-2000-Thompson-ApJ...534..239M,Thompson-2000-Madau-ApJ...538..105T,
Meszaros-2001-Ramirez-Ruiz-ApJ...554..660M,Beloborodov-2002-ApJ...565..808B,
Beloborodov-2005-ApJ...627..346B,Beloborodov-2014-Hascoet-ApJ...788...36B}
before the front of the ejecta catching up with it.
The observed optical and X-ray afterglow bump is due to the ejecta propagates into the $e^{\pm}$-rich medium shell.
Fifth, the VHE gamma-ray afterglow observed with H.E.S.S. is produced via the SSC process.

In this paper, we do not calculate the gamma-ray scattering by the medium for producing the soft gamma-ray pulse,
but focus on fitting the radio, optical and X-ray afterglow light curves
within this paradigm with the standard forward shock model,
in which the afterglow emission is attributed to the synchrotron radiation,
synchrotron-self-Compton (SSC) scattering of relativistic electrons accelerated via external forward shocks
(e.g., \citealp{Fan-2008-Piran-MNRAS.384.1483F,Ren-2020-Lin-ApJ...901L..26R}). Moreover, the effect of equal-arrival-time surface is also considered in the calculations \citep{Waxman-1997ApJ...491L..19W}.
The jet is assumed as a top-hat jet without lateral expansion \citep{Huang-1999-Dai-MNRAS.309..513H}.
The synchrotron self-absorption, Klein-Nishina effect, and the $\gamma\gamma$ annihilation effects are also considered
(e.g., \citealp{Gould-1967-Schreder-PhRv..155.1404G,Granot-1999-Piran-ApJ...527..236G,
Huang-2020-Liang-ApJ...903L..26H,Zhang-2020-Murase-arXiv201207796Z}).
We assume that the optical and X-ray bump since $t>700$ seconds
is due to the jet propagates into the $e^{\pm}$-rich medium,
which is assumed to be a homogeneous shell with uniform density.
Since the $e^{\pm}$-rich medium shell is resulted from the  $\gamma\gamma$ annihilation,
the dynamic evolution of the ejecta does not change before and after its propagation in the $e^{\pm}$-rich medium shell.
We define the $e^{\pm}$-enriched shell with three parameters,
$R_{\rm s}$, $R_{\rm e}$, and $k$, where $R_{\rm s}$ and $R_{\rm e}$ is the inner and outer boundary radius
of the $e^{\pm}$-enriched shell,
$k$ is the lepton number density ratio of $e^{\pm}$-rich medium shell to the normal medium.

We use a Markov Chain Monte Carlo (MCMC) algorithm
({\tt emcee}, \citealp{Foreman-Mackey-2013-Hogg-PASP..125..306F})
to fit the multiwavelength light curves of GRB~190829A afterglow. Our result is shown in the left panel of Figure~{\ref{multiwavelength}}.
One can observe that the light curves are well fitted with our model.
The model light curve in 15.5 GHz has a similar bump feature in the early stage staring at $\sim 700$ seconds,
but it has a long plateau up to $10^{5}$ seconds.
The model flux in the 1.3 GHz steady increases and peaks at $2\times 10^6$ seconds.

The obtained parameters of jet and their $1\sigma$ confidence levels are shown as following:
the isotropic kinetic energy ${\rm log_{10}}E_{\rm k, iso} ({\rm erg}) =51.01_{-0.33}^{+0.43}$,
the initial bulk Lorentz factor ${\rm log_{10}}\Gamma_{0}=1.55_{-0.33}^{+0.17}$,
the circum-burst medium number density ${\rm log_{10}}n_0({\rm cm}^{-3})=0.34_{-0.70}^{+0.56}$,
the electron energy fraction ${\rm log_{10}}\epsilon_e=-0.49_{-0.22}^{+0.46}$, the magnetic field energy fraction
${\rm log_{10}}\epsilon_B=-3.22_{-0.80}^{+1.21}$,
and the spectrum index of electrons $p=2.12_{-0.17}^{+0.08}$.
The parameters of the $e^{\pm}$-rich shell are
${\rm log_{10}}R_s ({\rm cm})=16.61_{-0.14}^{+0.30}$,
${\rm log_{10}}R_e({\rm cm})=16.81_{-0.15}^{+0.38}$,
and $k=6.87_{-2.55}^{+3.64}$.

Different from typical GRBs,
the jet of GRB~190829A is middle relativistic, with $\Gamma_0\sim 35$.
No jet break is observed until $t>5.5\times 10^{6}$ seconds when the afterglow are dimmer than the host galaxy.
Taking the jet break time $t_{\rm j}>5.5 \times 10^{6}$ seconds,
we calculate the jet half-opening angle with \citep{Frail-2001-Kulkarni-ApJ...562L..55F}
\begin{equation}
\begin{array}{ll}
\theta_{\rm j} &= 0.057(\frac{t_{\rm j}}{1\rm  day})^{3/8}(\frac{1+z}{2})^{-3/8}[\frac{E_{\rm \gamma, iso}}{10^{53}\rm  ergs}]^{-1/8} \\
                     & \times(\frac{\eta_{\gamma}}{0.2})^{1/8}(\frac{n}{0.1\rm cm^{-3}})^{1/8}~,
\end{array}
\end{equation}
where $E_{\rm \gamma,iso}$, $\eta_{\gamma}$
are the isotropic gamma-ray energy and radiative efficiency, respectively.
The circumburst medium density is taken as $n_0$.
The GRB efficiency is calculated as $\eta_{\gamma}=E_{\rm \gamma,iso}/(E_{\rm \gamma,iso}+E_{\rm k,iso})=17\%$ with $E_{\rm \gamma,iso}=2\times10^{50}$ erg \citep{Tsvetkova-2019-Golenetskii-GCN.25660....1T}. We obtain $\theta_j>1.0$ rad, indicating that the ejecta is almost isotropic.

\section{VHE Gamma-Ray Afterglows}
\label{VHE_afterglow}
H.E.S.S. detected the VHE gamma-rays of GRB~190829A afterglow with a confidence level of $5\sigma$
in the time interval from $t=\rm T_0+4h20m$ to $t=\rm T_0+7h54m$,
but the observed flux is not released
\citep{de_Naurois-2019-H.E.S.S._Collaboration-GCN.25566....1D}
\footnote{The H.E.S.S. observational data was published \citep{H.E.S.S. Collaboration} when our manuscript is under reviewing.
We also added the H.E.S.S. data in Figure~{\ref{SED}}.},
where $\rm T_0$ is the trigger time of {\em Fermi}/GBM.
Therefore, our above afterglow modeling does not take the VHE gamma-ray afterglow into account.
We examining whether our model calculation satisfies the observation with H.E.S.S., and presents
a further discussion on whether it can be detectable with the current and near-future telescopes.

Figure~{\ref{SED}} shows the 0.5~TeV and 0.2 - 4 TeV (H.E.S.S. energy band, \citealp{H.E.S.S. Collaboration}) lightcurves and
the SEDs at 1200 seconds (the peak time of the light curve) and $2\times 10^4$ seconds (in the time interval of H.E.S.S. observation for GRB~190829A).
The light curve is corrected for the EBL absorption \citep{Dominguez-2011-Primack-MNRAS.410.2556D}.
One can find that VHE afterglow is very bright during the ejecta propagates in the $e^\pm$-rich shell.
The afterglow light curve of GRB~190114C in the 0.5~TeV band is also shown in Figure~{\ref{SED}} for comparison.
One can find that flux of the VHE gamma-ray afterglow of GRB~190829A is comparable to GRB~190114C
at the epoch from $[2\times 10^3, 1\times 10^4]$ seconds.

The SEDs illustrates that the $\gamma\gamma$ annihilation effect
in the ejecta is considerable, especially in the early epoch. 
Taking the EBL absorption effects into account \citep{Dominguez-2011-Primack-MNRAS.410.2556D},
the VHE afterglow at this time epoch are convincingly detectable with H.E.S.S.,
MAGIC and CTA (Cherenkov Telescope Arrays),
but is not detectable with LHAASO (Large High Air Altitude Shower Observatory).
The SED at $t=\rm T_0+2 \times 10^4~s$ shows that the VHE emission is still can be detectable with H.E.S.S., which is consistent with the observations
\citep{de_Naurois-2019-H.E.S.S._Collaboration-GCN.25566....1D}.
It is also marginally detectable with MAGIC and convincingly detectable with CTA.

\section{Conclusions and Discussion}
\label{Conclusions}
We have revisited the multi-wavelength data of nearby LL-GRB~190829A
and presented a self-consistent paradigm for interpreting its features by assuming
an $e^{\pm}$-rich medium shell resulted from the interaction between the hard gamma-ray photons and
dense medium pre-accelerated by the prompt gamma-rays.
We show that the observed radio, optical, X-ray and VHE gamma-ray afterglows
are attributed to the emission from the synchrotron radiation and
the SSC process of the electron accelerated in the forward shocks.
The results of our fit to the multi-wavelength afterglow light curves
show that the ejecta of GRB~190829A is almost isotropic ($\theta_j>1.0$ rad) and
middle relativistic ($\Gamma_0\sim 35$).
The electron density of the $e^{\pm}$-rich medium shell is $\sim 15$ cm$^{-3}$,
about 7 times higher than the electron density of its normal surrounding medium.
Based on the model parameters derived in our analysis,
we calculate the VHE gamma-ray light curve and the SED at the peak time of the light curve.
It is found that its VHE gamma-ray emission is convincingly detectable
with H.E.S.S., MAGIC, and CTA at its peak time.

The dusty medium should be essential for interpreting the data of GRB~190829A.
The detection of associated SN~2019oyw with GRB~190829A confirms its progenitor as a massive star.
\cite{Liang-2007-Zhang-ApJ...670..565L} proposed that local LL-GRBs would be a
unique GRB population with low-luminosity, small beaming factor,
and large local event rate. The ejecta of GRB~190829A is middle relativistic and quasi isotropic,
being consistent with the features of the local LL-GRBs.
Our joint optical-X-ray spectral analysis reveals that the ambient medium of the GRB is extremely dusty.
By correcting $i$ band SN data with a host galaxy
extinction of $E_{\rm B-V}=0.757$ by assuming a Milky Way (MW) extinction law\footnote{The $E_{\rm B-V}$ value reported in \cite{Chand-2020-Banerjee-ApJ...898...42C} is 1.04 by
adopting a the Small Magellanic Cloud (SMC) extinction law.
Besides, the difference in time interval selection for joint spectral
fit may also lead to derive different extinction value.},
we find that the peak absolute magnitude of SN~2019oyw at $i$ band is $M_i=-18.3 \pm 0.01$~mag,
which is comparable to SN~2006aj ($M_i=-18.36\pm0.13$~mag) and SN~2010bh ($M_i=-18.58\pm0.08$~mag) \citep{Hu-2020-Castro-Tirado-arXiv200904021H}.

We collect the $A_{\rm V}$ and $N_{\rm H}$ values of the GRBs associated with SN
from literature, as reported in Table~{\ref{Tab2}}.
Figure~{\ref{N_H-A_v} shows $N_{\rm H}$ as a function of $A_{\rm V}$.
A sample of typical long GRBs taken from literature,
together with the typical dust-to-gas ratio
for the local group (LG) environment
$N_{\rm H}/A_{\rm V}=1.6\times10^{22}~{\rm cm}^{-2}\;{\rm mag}^{-1}$
\citep{Covino-2013-Melandri-MNRAS.432.1231C},
is also shown in Figure~{\ref{N_H-A_v}}.
One can observe that the $A_{\rm V}$ value of GRB~190829A
is the largest one among the SN-associated GRB sample,
although it is still not significantly distinct from the typical GRBs.
It closes to the dust-to-gas ratio for the LG environment,
but is above the ratio as most long GRBs.

It was proposed that the interaction between the prompt gamma-ray pulse and
the medium not only can pre-accelerate the ambient medium to a high Lorentz factor,
but also accompany by $e^{\pm}$ loading via the pair production process (e.g., \citealp{Thompson-2000-Madau-ApJ...538..105T,
Madau-2000-Thompson-ApJ...534..239M,Meszaros-2001-Ramirez-Ruiz-ApJ...554..660M}).
This may affect the GRB afterglow behaviors (e.g., \citealp{Beloborodov-2002-ApJ...565..808B}).
We suspect that the $e^{\pm}$-rich medium shell is resulted from the interaction
of the initial extremely hard gamma-ray photons of GRB~190829A with the dusty medium ($A_{\rm V}=2.33$).
As proposed by \cite{Shao-2007-Dai-ApJ...660.1319S},
the dust scattering of the gamma-ray photons may lead to softening echo emission.
This may naturally explain the observed soft gamma-ray pulse with a time decay
of 50~seconds with respect to the first gamma-ray pulse.
We compare them in the inset of Figure \ref{Prompt_LC} by normalizing their flux levels and
making alignment with respect to the starting time of the two pulses.
It is find that their initial rising parts are similar,
but the second pulse is more symmetric than the first pulse with
a FWHM about twice of the first pulse.
It is possible that the symmetric feature may be due to
echo emission during scattering.

Because the $e^{\pm}$-rich medium shell is formed at 50 second post the first pulse,
the observed time decay of afterglow bump is $\sim 10^3$ seconds.
Our fit indicates that the $e^{\pm}$-rich medium shell is at a region of $R=(4.07-6.46)\times 10^{16}~\rm cm$.
This is generally consistent with the forward shock region for the GRB afterglows \citep{Mu-2016-Lin-ApJ...831..111M}.
The time delay for the ejecta catching up with the $e^{\pm}$-rich medium shell is
estimated as $\Delta t= R/c\Gamma_0^2\backsimeq 1.1\times 10^3$  seconds
without considering the deceleration of the ejecta\footnote{The ejecta may move in a cavity since the medium is pre-accelerated by the prompt gamma-rays \citep{Beloborodov-2002-ApJ...565..808B}.}.
This is consistent with the observation.

\acknowledgments
We acknowledge the use of the public data from the {\em Swift} data archive and the UK {\em Swift} Science Data Center. This research has made use of the CTA instrument response functions provided by the CTA Consortium and Observatory,
see \url{http://www.cta-observatory.org/science/cta-performance/}
(version prod3b-v2) for more details.
This work is supported by the National Natural Science Foundation of China
(Grant No.11533003, 11851304, and U1731239) and Guangxi Science Foundation (grant No. 2017AD22006).

\clearpage


\clearpage

%

\begin{table}[h]
\tabletypesize{\footnotesize}
\tablewidth{0pt}
\caption{Properties of the GRB-SN Samples with Multicolor Light Curves}
\centering
\begin{tabular}{llllll}
\hline\hline
  GRB/SN       &  Redshift   &    $A_{\rm V}^{\rm host}(\rm mag)^{a}$   &   $N_{\rm H}(10^{21}~{\rm cm})^{d}$   &   $T$(s)  \\
\hline
  980425/1998bw      &    $0.0085$    &   $0.17\pm0.02$   &   $-$                   &     $-$   \\
\hline
  030329/2003dh      &    $0.16867$   &   $0.39\pm0.15$   &   $-$                   &     $-$ \\
\hline
  050525A/2005nc     &    $0.606$     &   $0.36\pm0.05$   &   $5.9^{+3.6}_{-2.7}$   &     $\rm T_{0}+151542$  \\
\hline
  060218/2006aj      &    $0.03342$   &   $0.13\pm0.01$   &   $3.1^{+0.7}_{-0.6}$   &     $\rm T_{0}+7028$  \\
\hline
  081007/2008hw      &    $0.5295$    &   $0.31\pm0.25$   &   $7.7^{+1.5}_{-1.4}$   &     $\rm T_{0}+8027$   \\
\hline
  091127/2009nz      &    $0.49044$   &   $0.17\pm0.15$   &   $1.1^{+0.6}_{-0.6}$   &     $\rm T_{0}+13229$   \\
\hline
  100316D/2010bh     &    $0.0592$    &   $0.43\pm0.03$   &   $20^{+16}_{-11}$      &     $\rm T_{0}+133397$  \\
\hline
  101219B/2010ma     &    $0.55185$   &    $<0.1$         &   $0.7^{+0.6}_{-0.5}$   &     $\rm T_{0}+362754$   \\
\hline
  111209A/2011kl     &    $0.677$     &   $0.3\pm1.5$     &   $2.7^{+0.8}_{-0.8}$   &     $\rm T_{0}+267284$   \\
\hline
  130427A/2013c      &    $0.3399$    &   $0.13\pm0.06$   &   $1.22^{+0.18}_{-0.18}$ &    $\rm T_{0}+712132$   \\
\hline
  130702A/2013dx     &    $0.677$     &   $0.3\pm0.07$    &   $1.4^{+0.31}_{-0.3}$   &    $\rm T_{0}+1478125$  \\
\hline
  130831A/2013fu     &    $0.4791$    &   $0.06\pm0.04$   &   $0.2^{+3.88}_{-0.2}$  &     $\rm T_{0}+251766$   \\
\hline
  140606B/iPTF14bfu  &    $0.384$     &   $0.47\pm0.41^{b}$   &   $6^{+8}_{-5}$       &   $\rm T_{0}+187980$  \\
\hline
  171205A/2017iuk    &    $0.0368$    &   $0.155^{c}$         &   $1.2^{+0.8}_{-0.7}$   &  $\rm T_{0}+749534$   \\
\hline
  190829A/2019oyw    &    $0.0785$    &   $2.33$          &   $5.79^{+0.53}_{-0.53}$   &   $\rm T_{0}+(3-4)\times 10^4$   \\
\hline
\end{tabular}

\tablenotetext{a} {The values of $A_{\rm V}$, redshift $z$ from GRB~980425 to GRB~130831A are taken from \cite{Li-2018-Wang-ApJS..234...26L}, except for 140606B/iPTF14bfu from \cite{Cano-2015-de_Ugarte_Postigo-MNRAS.452.1535C}$^{b}$ and 171205A/2017iuk from \cite{Suzuki-2019-Maeda-ApJ...870...38S}$^{c}$.}
\tablenotetext{d}{Taken from the XRT Catalogue entry of {\em Swift} website and $T$ is the corresponding time for extracting the XRT spectrum.}
\label{Tab2}
\end{table}

\clearpage
\begin{figure}[htbp]
\centering
\includegraphics[angle=0,width=0.5\textwidth]{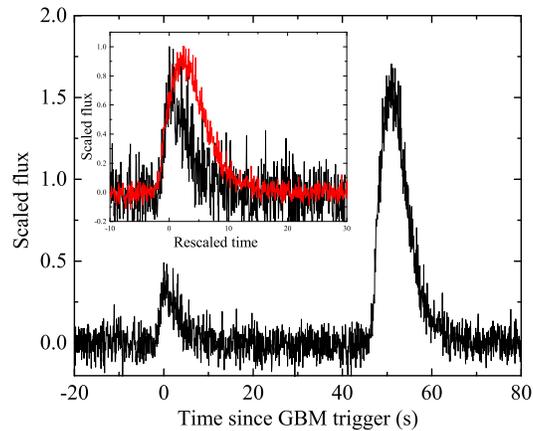}
\caption{Prompt gamma-ray light curve of GRB~190829A observed with {\em Fermi}/GBM.
The inset shows the comparison of the two gamma-ray pulses by re-scaling their flux level and making alignment of their beginning times.
}\label{Prompt_LC}
\end{figure}

\begin{figure}[htbp]
 \centering
\includegraphics[angle=0,width=0.4\textwidth]{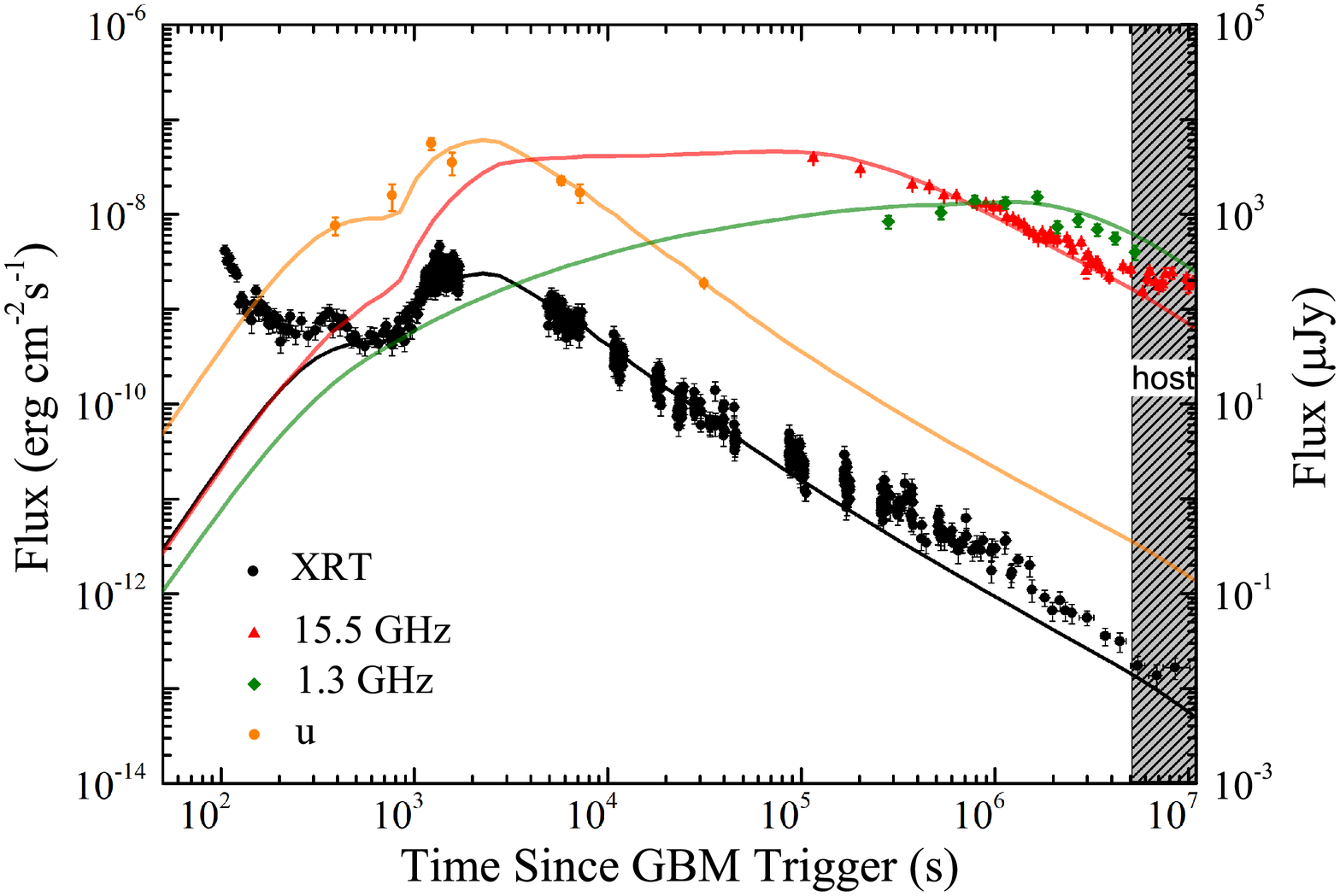}
\includegraphics[angle=0,width=0.4\textwidth]{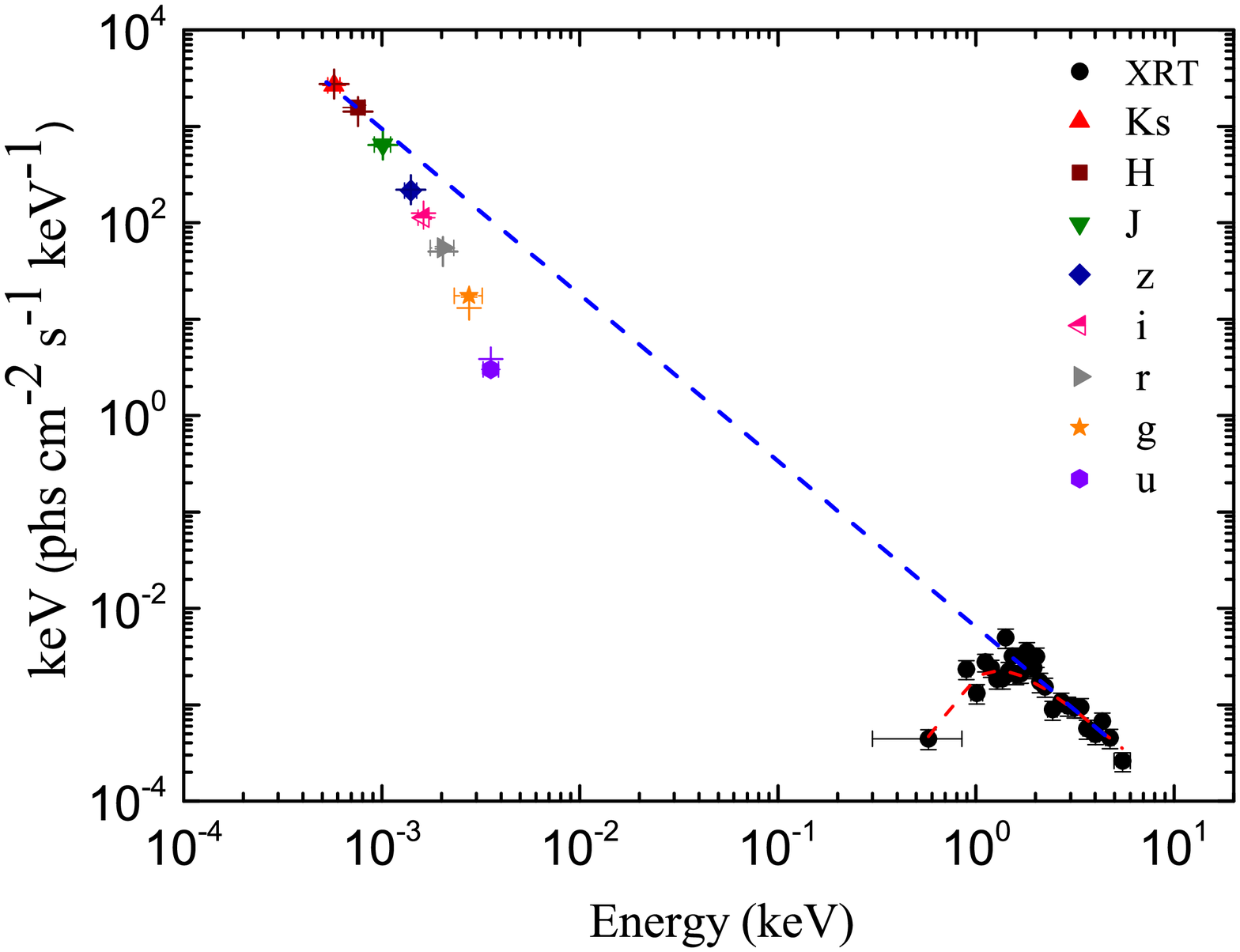}
\caption{{\em Left panel---} Multiwavelength afterglow lightcurves (dots) of GRB~190829A
along with our fits with the forward shock model (lines). The shaded region marks the
afterglows are dimmer than the emission of the host galaxy.
{\em Right panel---} Joint optical-X-ray afterglow spectrum (dots) observed
in the time interval of $[3,4]\times10^{4}~\rm s$ of GRB~190829A along with
our fit with a single power-law function (dashed lines). The optical data are
extinction-corrected for our Galaxy only. Extinction and HI absorbtion of both
our Galaxy and the GRB host galaxy are considered in our spectral fit.}
\label{multiwavelength}
\end{figure}

\clearpage

\begin{figure}[htbp]
\centering
\includegraphics[angle=0,width=0.32\textwidth]{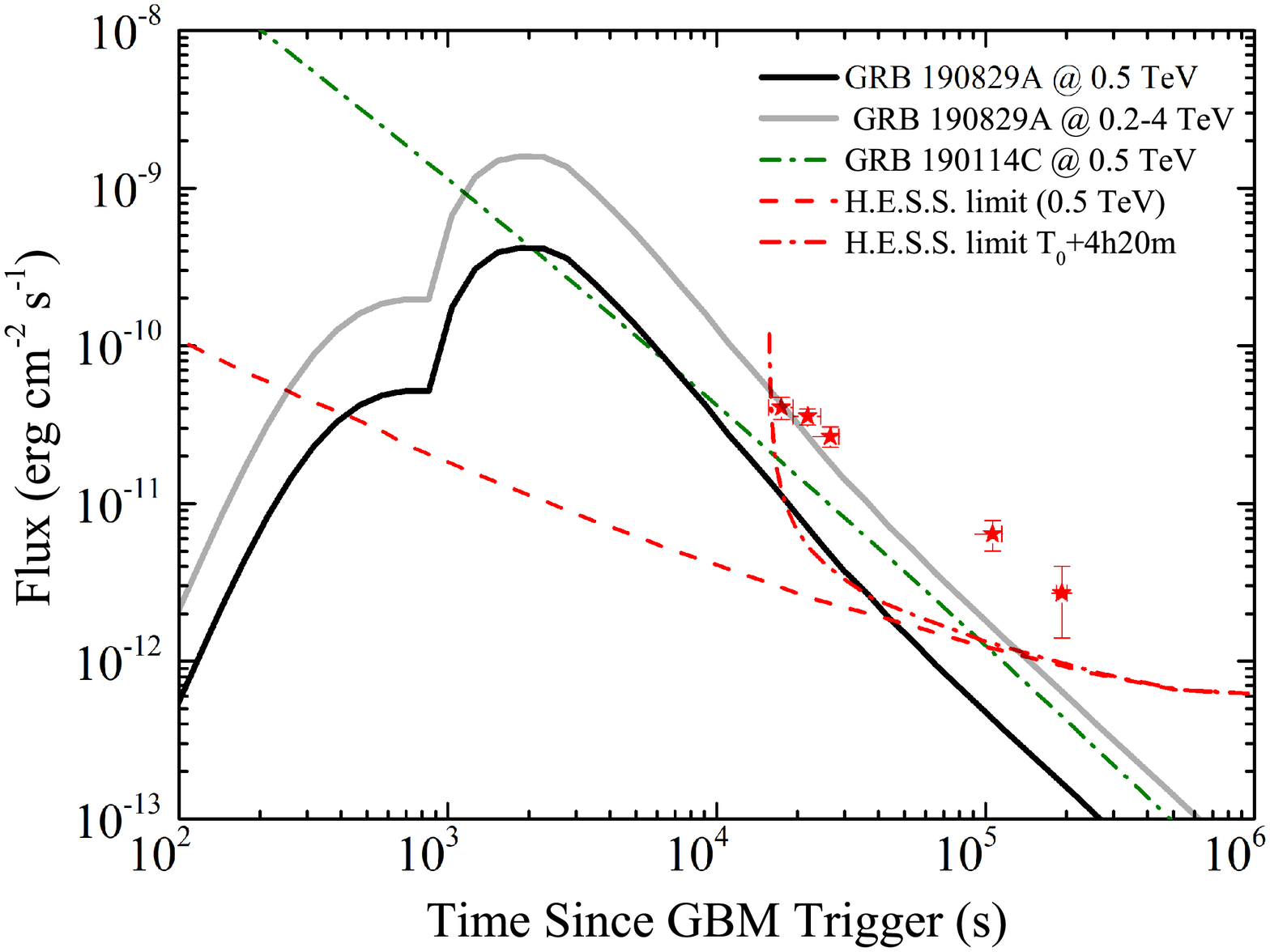}
\includegraphics[angle=0,width=0.32\textwidth]{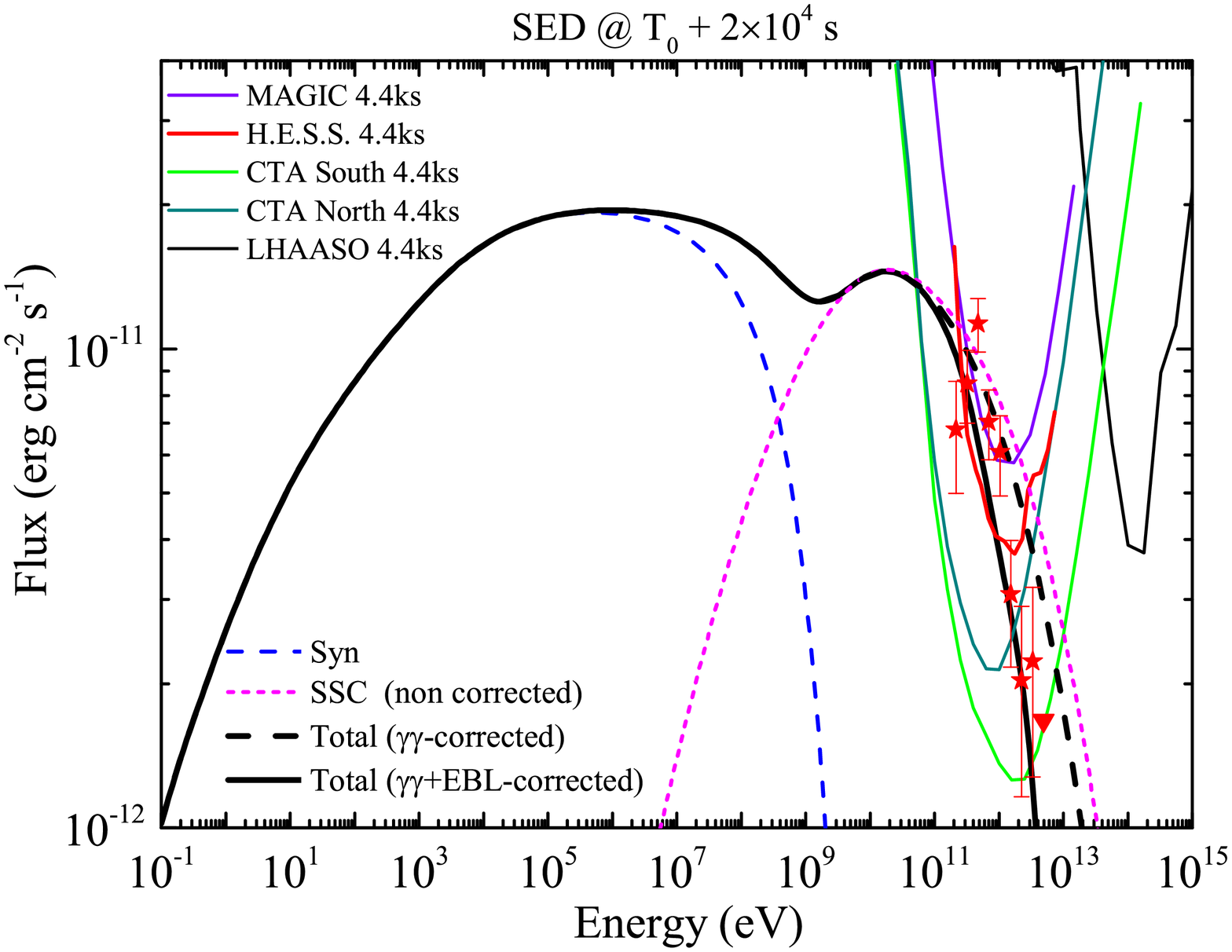}
\includegraphics[angle=0,width=0.32\textwidth]{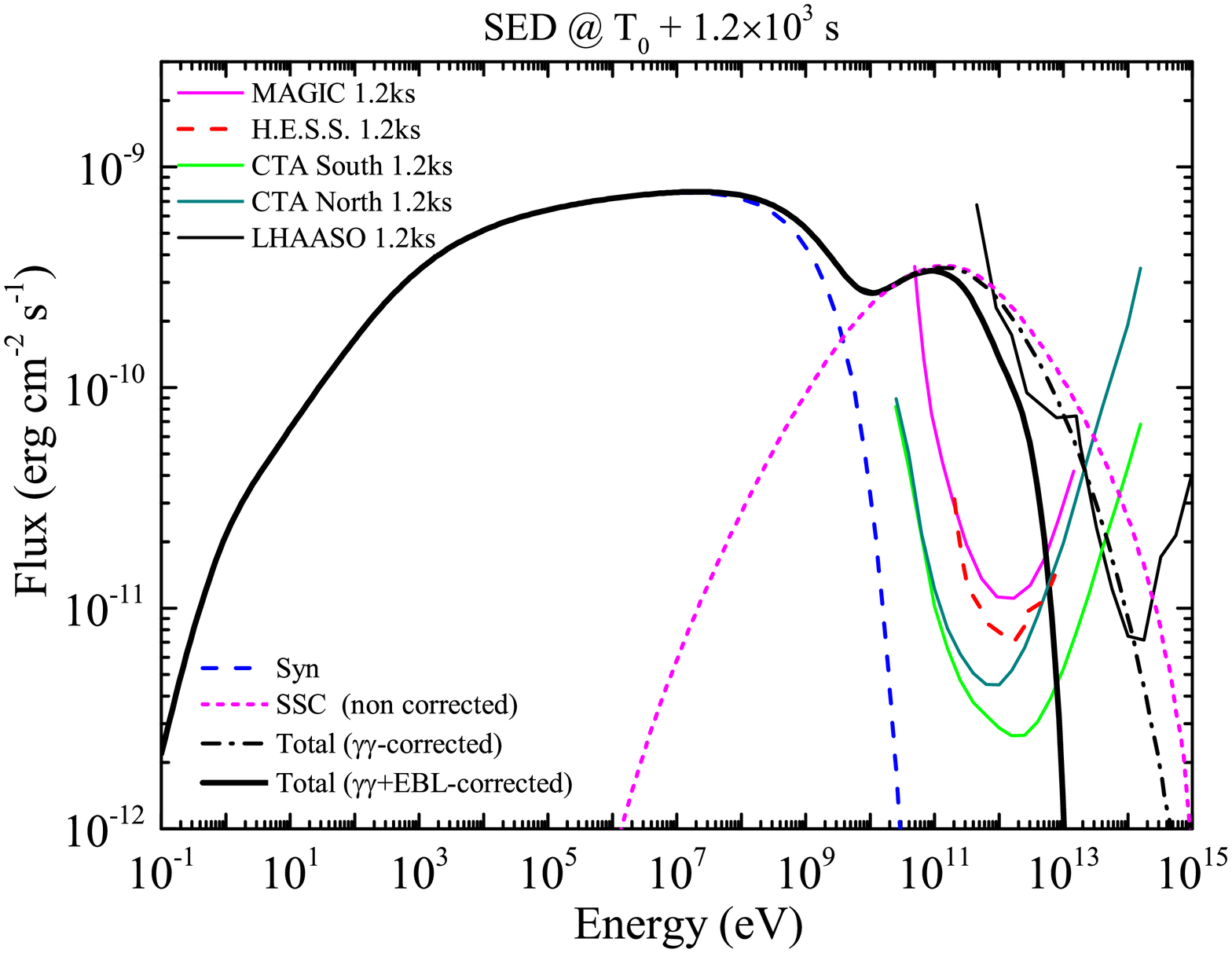}
\caption{{\em Left panel---} The theoretical lightcurves at 0.5~TeV (black solid line) and in the H.E.S.S. energy band (0.2 - 4~TeV, gray solid line)
of GRB~190829A. The H.E.S.S. data taken from \citep{H.E.S.S. Collaboration} are shown with red stars. The theoretical 0.5 TeV lightcurve of GRB 190114C (green dash-dotted line) is also shown for comparison. The $\gamma\gamma$ annihilation in the ejecta and EBL absorption are corrected.
The sensitivity curves at 0.5~TeV of H.E.S.S.
staring from the {\em Fermi}/GBM trigger time $\rm T_0$ and starting from $\rm T_0+4h20m$ are
shown in red dashed line and red dash-dotted line, respectively.
{\em Middle and Right panels---} The numerical SEDs of GRB~190829A at $2\times 10^4$~s and 1200~s, respectively. The H.E.S.S. data observed in the time interval [4.5, 7.9] hours taken from \citep{H.E.S.S. Collaboration} are shown with red stars. The sensitivity curves scaled to an observational times of $2\times 10^4$~s and 1200~s are also shown.
The sensitivity curves are adopted from \cite{Bai-2019-Bi-arXiv190502773B} and
\url{https://www.cta-observatory.org/science/cta-performance/}.
}\label{SED}
\end{figure}

\clearpage

\begin{figure}[htbp]
\centering
\includegraphics[angle=0,width=0.8\textwidth]{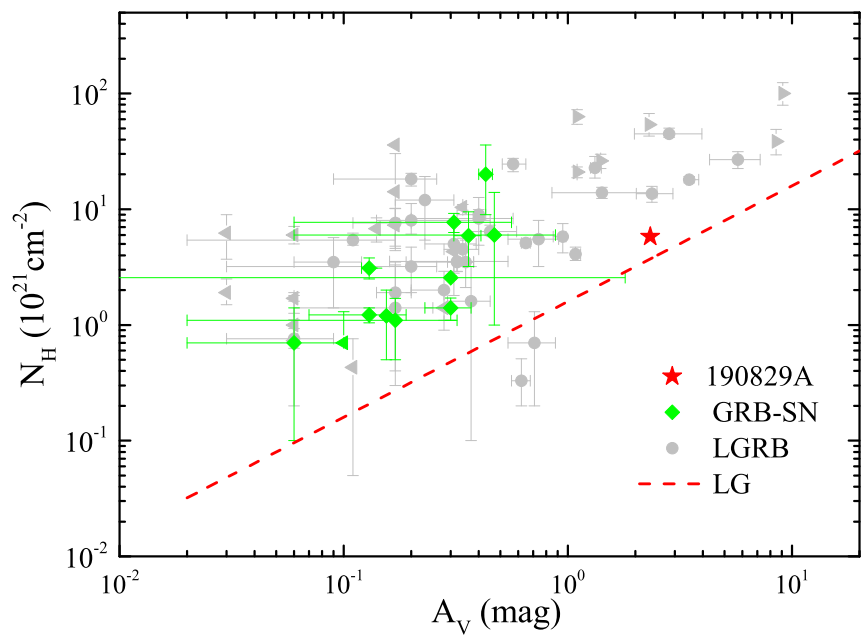}
\caption{$N_{\rm H}$ as a function of $A_{\rm V}$ of GRBs associated with an associated SNe (green diamond dots) in comparison with typical long GRBs (grey dots and triangles) from \cite{Covino-2013-Melandri-MNRAS.432.1231C}. Triangle represents the upper or lower limit. GRB~190829A-SN~2019oyw is marked as red star. The red dashed line shows typical dust-to-gas ratio for the local group (LG) environment \citep{Covino-2013-Melandri-MNRAS.432.1231C}.
}\label{N_H-A_v}
\end{figure}

\clearpage

\end{document}